\documentclass[conference,compsoc]{IEEEtran}
\ifCLASSOPTIONcompsoc
  \usepackage[nocompress]{cite}
\else
  \usepackage{cite}
\fi
\ifCLASSINFOpdf
\else
\fi

\usepackage{float} 
\usepackage{soul}

\usepackage{graphicx}
\begin{document}
%
\title{Deep Reinforcement Learning Strategies in Finance: Insights into Asset Holding, Trading Behavior, and Purchase Diversity\\
\footnotesize  Regular Research Paper (CSCE-ICAI'24)}


\author{\IEEEauthorblockN{Alireza Mohammadshafie}
\IEEEauthorblockA{Computer Science and Engineering\\
University of North Texas\\
Denton,Texas, USA \\
Email:Alirezamohammadshafie@my.unt.edu }

\and
\IEEEauthorblockN{ Akram Mirzaeinia}
\IEEEauthorblockA{\textit{Independent Researcher} \\
\textit{Not Affiliated(Independent) }\\
Ajman, United Arab Emirates\\
Email: mirzaei.kad@gmail.com}
\and
\IEEEauthorblockN{ Haseebullah Jumakhan}
\IEEEauthorblockA{\textit{Artificial Intelligence Research Center (AIRC)} \\
\textit{Ajman University}\\
Ajman, United Arab Emirates \\
0009-0005-5461-5531}
\and
\IEEEauthorblockN{Amir Mirzaeinia}
\IEEEauthorblockA{Computer Science and Engineering\\
University of North Texas\\
Denton Texas, USA\\
Email: Amir.Mirzaeinia@unt.edu}

}


%


\maketitle

\begin{abstract}

Recent deep reinforcement learning (DRL) methods in finance show promising outcomes. However, there is limited research examining the behavior of these DRL algorithms. This paper aims to investigate their tendencies towards holding or trading financial assets as well as purchase diversity. By analyzing their trading behaviors, we provide insights into the decision-making processes of DRL models in finance applications. Our findings reveal that each DRL algorithm exhibits unique trading patterns and strategies, with A2C emerging as the top performer in terms of cumulative rewards. While PPO and SAC engage in significant trades with a limited number of stocks, DDPG and TD3 adopt a more balanced approach. Furthermore, SAC and PPO tend to hold positions for shorter durations, whereas DDPG, A2C, and TD3 display a propensity to remain stationary for extended periods.
\end{abstract}


%
\IEEEpeerreviewmaketitle

\section{Introduction}
Artificial intelligence (AI), particularly in the domain of machine learning (ML), has been progressively revolutionizing enterprises through its ability to absorb information and adapt based on inputs. One prominent sub-field within machine learning (ML), reinforcement learning (RL), has garnered acclaim for its capability to enable agents to develop optimal decision-making strategies via interaction with their surroundings. Reinforcement learning algorithms, inspired by behavioral psychology, gain knowledge through a process of experimenting and receiving feedback in the form of rewards or penalties based on their actions. The iterative nature of this approach allows reinforcement learning agents to gradually improve their decision-making abilities and achieve optimal outcomes. The objective of this paper is to investigate the behavior and decision-making processes of DRL algorithms in financial trading, with a focus on their tendencies towards holding or trading assets and their purchase diversity.

An improved form of reinforcement learning (RL) called deep reinforcement learning (DRL) makes use of deep neural networks to analyze large-scale input spaces efficiently. DRL algorithms utilize the integration of deep learning techniques and RL to efficiently tackle complex real-world problems that were previously unachievable using traditional RL methods. One noteworthy characteristic of DRL algorithms is their ability to autonomously gain knowledge from unprocessed sensory input, including sensor data and photographs, without the need for domain-specific expertise or pre-designed features.

Classifying an algorithm as DRL relies on various crucial variables. Firstly, it must employ deep neural networks as function approximators to accurately represent the value functions or policies. Additionally, it should utilize reinforcement learning (RL) approaches, such as value iteration or policy gradients, to optimize the behavior of the agent by incorporating feedback from the environment. Finally, it must exhibit the capacity to acquire knowledge directly from unprocessed sensory information, allowing it to make generalizations across a wide range of complex and multidimensional facts.

The convergence of DRL (Deep Reinforcement Learning) and finance presents a significant opportunity to transform the process of making trading and investing choices. The complexity, volatility, and uncertainty of financial markets make them a perfect environment for the application of Deep Reinforcement Learning (DRL) algorithms. These algorithms have the capacity to reveal concealed patterns, take advantage of inefficiencies, and optimize trading strategies in real-time, thus improving market efficiency and maximizing returns for investors.

Within the domain of finance, decision-making is centered on a variety of tactics, spanning from long-term investment to short-term trading. To successfully implement any strategy, one must possess a sophisticated comprehension of market dynamics, skillful risk management, and the capability to adjust to evolving circumstances. Deep reinforcement learning (DRL) algorithms, similar to human traders, have the ability to determine whether to retain assets for an extended period or engage in frequent trading in the short term. Furthermore, their ability to navigate complex financial data in multiple dimensions allows them to discover detailed patterns and connections that human traders may not be able to find.

Additionally, spreading investment capital across different assets or securities are known to be a common approach to reduce the risk. Instead of putting all funds into a single asset or market, traders diversify their portfolios by investing in a variety of assets, such as stocks, bonds, commodities, and currencies, as well as across different sectors and geographical regions. The primary goal of trading diversification is to minimize the impact of adverse events or fluctuations in any single asset or market on the overall portfolio performance. By holding a diversified portfolio, traders aim to achieve a balance between risk and return, potentially improving the risk-adjusted returns of their investments. Diversification can take several forms such as asset allocation, sector diversification, geographical diversification, instrument diversification, time diversification. While diversification can help mitigate risk, it is important to note that it does not eliminate all risk, especially systemic risks that affect entire markets or economies. Additionally, over-diversification can dilute potential returns, so finding the right balance between diversification and concentration is crucial for portfolio management. In this paper, we additionally examine the behaviour of DRL-trained agents to diversify their trading strategies.

Previous research has investigated the application of various DRL algorithms in finance, focusing on their performance and potential to outperform traditional trading strategies employed by human actors \cite{montazeri2023scnnFinrl,montazeri2023scnnFinrl_shuffled}.This article aims to investigate the performance of different Deep Reinforcement Learning (DRL) algorithms within the field of finance.  Through analyzing their performance in diverse trading and investment situations, our objective is to acquire knowledge about their capabilities, constraints, and appropriateness for different financial uses. In addition, we analyze how these DRL algorithms make the crucial decision of whether to hold or trade assets as well as asset diversification, providing insight into their decision-making processes and the consequences for financial decision-makers. By conducting this investigation, our objective is to enhance comprehension of the impact of DRL on the evolution of the financial sector.
\section{Methodology}
\subsection{Data Source}
Yahoo finance is one of the top available financial data sources and a comprehensive financial platform that offers a wide range of tools, resources, and information for investors, traders, and financial professionals. Serving as one of the leading financial portals, Yahoo Finance provides users with access to real-time market data, news, analysis, research tools, and investment resources. Yahoo finance offers some key features and offerings as following.

Market Data: Yahoo Finance provides users with access to real-time and historical data on stocks, bonds, mutual funds, exchange-traded funds (ETFs), indices, currencies, commodities, and other financial instruments. Users can track prices, view charts, and analyze performance metrics for individual securities or market indices.

News and Analysis: Yahoo Finance offers a comprehensive selection of financial news articles, analysis, commentary, and insights from leading financial journalists and contributors. Users can stay informed about market trends, economic developments, corporate earnings, and other relevant news impacting the financial markets.
Portfolio Management: Yahoo Finance allows users to create and manage personalized investment portfolios, track holdings, monitor performance, and analyze portfolio metrics. Users can input their investment transactions, view portfolio allocation, and assess risk exposure to make informed investment decisions.

Screeners and Research Tools: Yahoo Finance offers powerful screening tools and research resources for identifying investment opportunities based on specific criteria, such as market capitalization, sector, industry, valuation metrics, and performance indicators. Users can filter stocks, ETFs, and mutual funds based on their investment preferences and objectives.
Educational Resources: Yahoo Finance provides educational articles, tutorials, videos, and guides to help users enhance their financial literacy, understand investment concepts, and improve their investment strategies. Topics cover a wide range of subjects, including investing basics, portfolio management, risk management, and market analysis.

Community and Social Features: Yahoo Finance fosters a vibrant community of investors and traders through its discussion forums, message boards, and social networking features. Users can engage with other members, share investment ideas, discuss market trends, and exchange insights and opinions.

Overall, Yahoo Finance serves as a comprehensive and user-friendly platform for individuals seeking to stay informed about the financial markets, manage their investments, conduct research, and engage with a community of like-minded investors. Its wide range of features and resources make it a valuable tool for investors of all levels of expertise.

 The financial data used in this analysis was obtained from Yahoo Finance. The data covers hourly intervals from March 4, 2022, to March 1, 2024. The dataset was partitioned into a training period spanning from March 4, 2022, to December 1, 2023, and a testing period spanning from December 1, 2023, to March 1, 2024. This dataset contains price data for each of the thirty companies featured in the Dow Jones Industrial Average listed in table \ref{tab:comp_sectors}. The data includes the opening, low, high, and closing values, and is recorded on an hourly basis.

\begin{table}[]
\centering
\caption{List of compnaies in our dataset and their sectors}
\label{tab:comp_sectors}
\begin{tabular}{|l|l|}
\hline
Sector name                                                       & Companies                                                                                                                                                                                   \\ \hline
Technology                                                        & \begin{tabular}[c]{@{}l@{}}The Home Depot Inc. (HD)\\ McDonald's Corporation (MCD)\\ Nike, Inc. (NKE) \\ The Walt Disney Company (DIS)\\ Walmart Inc. (WMT)\end{tabular}                    \\ \hline
\begin{tabular}[c]{@{}l@{}}Consumer \\ Staples\end{tabular}       & \begin{tabular}[c]{@{}l@{}}The Coca-Cola Company (KO)\\ Procter \& Gamble Co. (PG)\\ Walgreens Boots Alliance, Inc. (WBA)\end{tabular}                                                      \\ \hline
Financials                                                        & \begin{tabular}[c]{@{}l@{}}American Express Company (AXP)\\ Goldman Sachs Group, Inc. (GS)\\ JPMorgan Chase \& Co. (JPM)\\ Visa Inc. (V)\\ The Travelers Companies, Inc. (TRV)\end{tabular} \\ \hline
Healthcare                                                        & \begin{tabular}[c]{@{}l@{}}Amgen Inc. (AMGN)\\ Johnson \& Johnson (JNJ)\\ Merck \& Co.\\ Inc. (MRK)\\ UnitedHealth Group Incorporated (UNH)\end{tabular}                                    \\ \hline
\begin{tabular}[c]{@{}l@{}}Consumer\\  Discretionary\end{tabular} & \begin{tabular}[c]{@{}l@{}}The Home Depot, Inc. (HD)\\ McDonald's Corporation (MCD)\\ Nike, Inc. (NKE)\\ The Walt Disney Company (DIS)\\ Walmart Inc. (WMT)\end{tabular}                    \\ \hline
Industrials                                                       & \begin{tabular}[c]{@{}l@{}}The Boeing Company (BA)\\ Caterpillar Inc. (CAT)\\ Honeywell International Inc. (HON)\\ 3M Company (MMM)\end{tabular}                                            \\ \hline
Energy                                                            & Chevron Corporation (CVX)                                                                                                                                                                   \\ \hline
Materials                                                         & Dow Inc. (DOW)                                                                                                                                                                              \\ \hline
Telecom                                                           & Verizon Communications Inc. (VZ)                                                                                                                                                            \\ \hline
\end{tabular}
\end{table}

\subsection{Utilized Indicators}
To guide our trading decisions, we employed an extensive range of technical indicators generated from the financial data listed in table \ref{tab:Utilized_Indicators}. The first indicators used is the Volatility Index (VIX) which is a measure of market volatility and investor sentiment in the stock market. The VIX is calculated by the Chicago Board Options Exchange (CBOE) based on the prices of options on the S\&P 500 index. Options are financial derivatives that give investors the right, but not the obligation, to buy or sell an underlying asset (in this case, the S\&P 500 index) at a predetermined price (strike price) within a specific period of time.

The second indicator is Moving Average Convergence Divergence (MACD), which is s a popular technical indicator used in financial analysis to identify changes in the strength, direction, momentum, and duration of a trend in a security's price.

Bollinger Bands are a popular technical analysis tool used by traders to analyze price volatility and potential price breakouts in financial markets. They consist of three lines of upper (BOLL\_UB), middle and lower band (BOLL\_LB).

Relative Strength Index (RSI\_30) which is,  is a momentum oscillator that measures the speed and change of price movements in financial markets. It is a popular technical analysis tool used by traders and investors to identify overbought or oversold conditions in a security's price and to gauge the strength of a trend. The RSI is calculated based on the following formula:
\begin{equation}
    RSI = 100 - \frac{100}{1+\frac{Average Gain}{Average Loss}}
\end{equation}
where average gain is the sum of gains over the specified period over Number of periods and average loss is the sum of losses over the specified period over Number of periods. The RSI typically uses a 14-period timeframe for its calculations, but traders may adjust this parameter based on their preferences and the characteristics of the security being analyzed.

The RSI ranges from 0 to 100 and is plotted as a line graph. The RSI value fluctuates between these two extremes, with readings above 70 typically considered overbought and readings below 30 considered oversold. Overbought conditions may suggest that the security is due for a potential reversal or correction, while oversold conditions may indicate a possible buying opportunity.

Commodity Channel Index (CCI\_30), is a versatile technical analysis indicator used to identify potential trends, overbought or oversold conditions, and price reversals in financial markets, particularly in commodities trading. It is developed by Donald Lambert in the late 1970 and it measures the relationship between an asset's current price, its moving average, and its typical price range. CCI can be calculated as following formula 
\begin{equation}
    CCI = \frac{Typical Price - SMA}{0.015*MeanDeviation}
\end{equation}
where typical price of an asset is the average of the high, low, and closing prices over a specified period. Simple Moving Average (SMA) calculate the n-period simple moving average of the typical price, where n is the number of periods used for the CCI calculation. Additionally, mean deviation calculate the mean deviation of each typical price from the SMA over the specified period. The mean deviation is the average of the absolute differences between each typical price and the SMA.

Directional Movement Index (DX\_30) is a technical indicator used to assess the strength and direction of a trend in financial markets. Developed by J. Welles Wilder Jr., the DMI consists of three lines: the Positive Directional Indicator (+DI) which measures the strength of upward price movement, the Negative Directional Indicator (-DI) which measures the strength of downward price movement, and the Average Directional Index (ADX), which measures the overall strength of the trend, regardless of its direction.  $\pm DI$ and $DX$ can be calculated in following equations.
\begin{equation}
    +DI = \frac{Smoothed +DM}{ATR}\times 100
\end{equation}
\begin{equation}
    -DI = \frac{Smoothed - DM}{ATR}\times 100
\end{equation}
\begin{equation}
    DX = \frac{|+DI- -DI|}{|+DI+ -DI|}\times 100
\end{equation}
where +DM is current high minue previous high. -DM is previous low minus current low and moothed +/-DM $= \sum_{t=1}^{14} DM - (\frac{\sum_{t=1}^{14} DM}{14})+CDM$. CDM is current DM and ATR is average true range.

We also have 30-period and 60-period Simple Moving Averages (SMA) which is average price of a security over a specified number of periods (30 and 60 hours). It is a lagging indicator, meaning it is based on past prices and does not predict future price movements. $SMA_{30}^i$ and $SMA_{60}^i$ can be calculated in following equations \ref{eq:SMA30} and \ref{eq:SMA60}.
\begin{equation}
    SMA_{30}^i = \frac{P_{i-1}+P_{i-2}+...+P_{i-30}}{30}
    \label{eq:SMA30}
\end{equation}
\begin{equation}
    SMA_{60}^i = \frac{P_{i-1}+P_{i-2}+...+P_{i-60}}{60}
    \label{eq:SMA60}
\end{equation}
where $P_{i+k}$ in these equations is the price of an asset at period k.
In addition, we integrated a turbulence indicator to measure market volatility and uncertainty.

\begin{table}[!t]
\caption{Utilized Indicators}
\label{tab:Utilized_Indicators}
\resizebox{\columnwidth}{!}{%
\begin{tabular}{|l|l|}
\hline
  & Indicators                                 \\ \hline
1 & Volatility Index (VIX)                     \\ \hline
2 & Moving Average Convergence Divergence (MACD)                     \\ \hline
3 & Bollinger Bands (BOLL\_UB and BOLL\_LB)    \\ \hline
4 & Relative Strength Index (RSI\_30)          \\ \hline
5 & Commodity Channel Index (CCI\_30)(CCI\_30) \\ \hline
6 & Directional Movement Index (DX\_30)        \\ \hline
7 & Close 30 Simple Moving Average (MSA)        \\ \hline
8 & Close 60 Simple Moving Average (MSA)        \\ \hline
\end{tabular}%
}
\end{table}

\subsection{Environment:}
Our experimentation is conducted using the FinRL environment \cite{li2021finrl}, which is a dedicated framework designed exclusively for the use of reinforcement learning (RL) in financial markets. FinRL provides a comprehensive set of tools specifically designed to address the distinct difficulties presented by financial time series data, including non-stationarity and high dimensionality. The platform offers a versatile and expandable setting for training reinforcement learning agents to make trading decisions across diverse market situations.

\subsection{Hyperparameters}
We adjusted the hyperparameters of our reinforcement learning models to maximize their performance on the specified task. The crucial hyperparameters consist of the number of time steps, which is set at 100,000, and it defines the length of each training session. By using this time step option, each agent goes through 33 complete episodes of training. At the onset of every episode, the agent commences with an initial capital of 1,000,000. The status space is composed of 301 dimensions (table \ref{tab:featurevectore}), which include a range of market indicators. The action space consists of three options: selling, buying, or holding stocks.

 \begin{table}[htbp]
\caption{Daily Feature Vector}
\begin{center}
\begin{tabular}{|l|l|}
\hline
Name                         & size \\ \hline
Amount                       & 1    \\ \hline
Price                        & 30   \\ \hline
Share held                   & 30   \\ \hline
Financial ratios (8 * 30)     & 240  \\ \hline
Total size of feature vector & 301  \\ \hline
\end{tabular}
\label{tab:featurevectore}
\end{center}
\end{table}

\subsection{Models Utilized}
We utilized a varied range of reinforcement learning algorithms to create and assess our trading strategies. The algorithms in question are Deep Deterministic Policy Gradient (DDPG) \cite{DDPG_lillicrap2015continuous}, Proximal Policy Optimization (PPO) \cite{ppo2017}, Twin Delayed DDPG (TD3) \cite{TD3_fujimoto2018addressing}, Soft Actor-Critic (SAC) \cite{SAC_haarnoja2018soft}, and Advantage Actor-Critic (A2C)\cite{A2C_mnih2016asynchronous}. These models have unique principles and learning mechanisms, leading to performance variations. Recent studies have also explored the use of convolutional neural networks (CNNs) in deep reinforcement learning for financial trading, demonstrating their ability to handle large, continuous action spaces \cite{montazeri2023scnnFinrl_shuffled,montazeri2023scnnFinrl}. However, we shall only focus on the models mentioned above to conduct an appropriate comparison of models in this study. DDPG employs an actor-critic framework with deterministic policies, whereas PPO concentrates on maximizing the surrogate objective function to provide reliable policy updates. TD3 incorporates two critics and delayed policy updates to enhance stability and improve sampling efficiency. SAC utilizes entropy regularization to promote exploration, while A2C utilizes several actors to parallelize policy updates and improve learning speed.

In the next Results section, we thoroughly examine the evaluation of these models and analyze their performance in trading scenarios.
\section{Results}
The results of our extensive experimental analysis  focuses on three key aspects: reward, purchase diversity, and the performance comparison between holder and trader algorithms. We examine the effectiveness of the reward mechanisms employed in our deep reinforcement learning (DRL) models, assessing their ability to incentivize desirable trading behaviors and optimize performance outcomes. Additionally, we investigate the diversity of asset purchases made by the DRL agents, exploring the breadth and distribution of investments across different financial instruments. Finally, we conduct a comparative analysis between two distinct trading strategies - holding and active trading - to evaluate their respective impact on portfolio performance and overall profitability. Through these comprehensive analyses, we aim to provide insights into the decision-making processes and efficacy of DRL models in financial applications.

\subsection{Reward}
In DRL models, rewards play a pivotal role in guiding the learning process. Rewards serve as feedback signals that inform the agent about the desirability of its actions in a given environment. By maximizing cumulative rewards over time, these algorithms learn to make decisions that lead to favorable outcomes. Rewards provide a clear objective for the agent, shaping its behavior and driving it towards optimal policies. In the absence of rewards or with sparse rewards, learning becomes challenging as the agent struggles to discern which actions contribute to long-term success. Thus, we carefully investigate the reward that is return by different DRL algorithms.

\begin{figure}[htbp]
    \centering
    \includegraphics[width=1\linewidth]{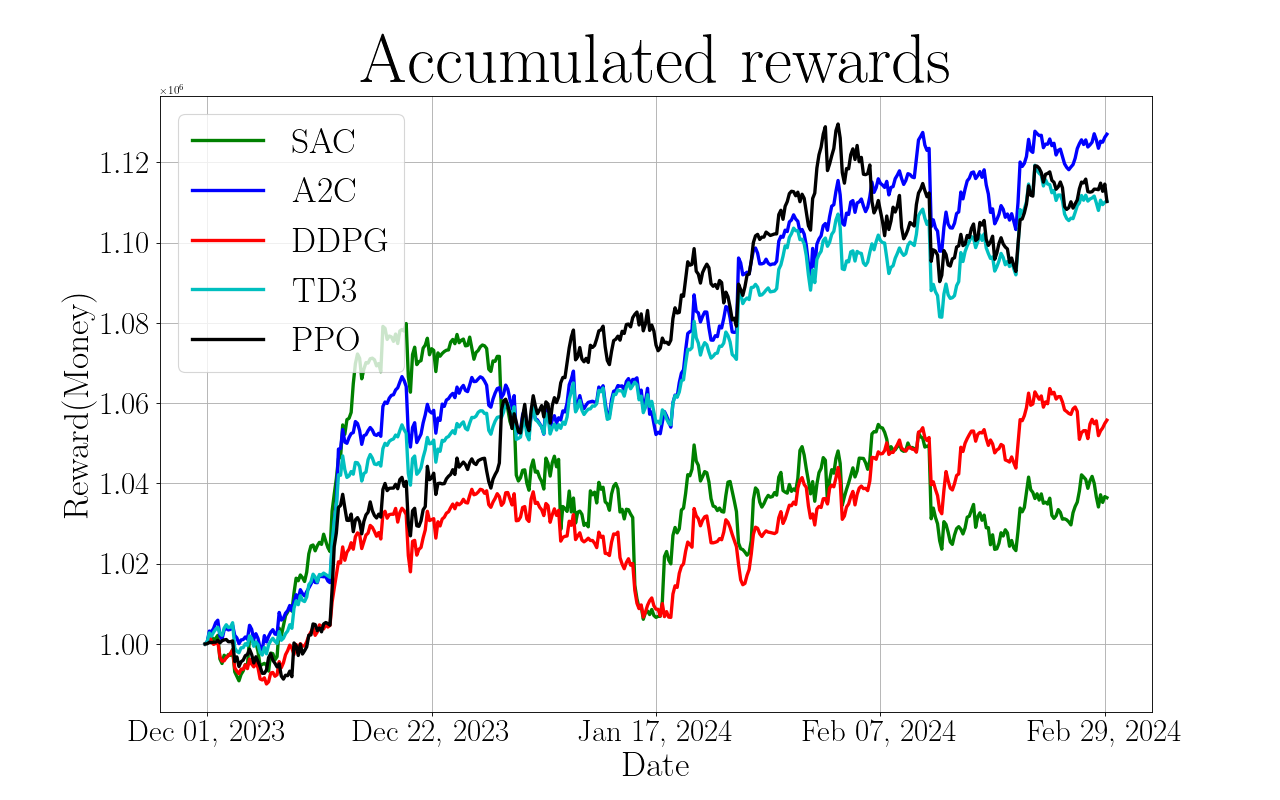}
    \caption{The temporal progression of accumulated rewards is analyzed throughout test data to reveal the performance dynamics of DDPG, PPO, TD3, SAC, and A2C models in real-world trading scenarios.}
     \label{fig:reward}
\end{figure}

The accumulative rewards plot (Figure \ref{fig:reward}) illustrates the effectiveness of each model's trading strategy in producing profits. Despite our initial expectations, which were in favor of SAC and PPO because to their success in other domains like training the humanoid Mujoco, A2C proved to be the top performer in terms of cumulative rewards. PPO and TD3 had somewhat lower performance, whilst DDPG and SAC lagged behind. This discrepancy underscores the intricacies of financial markets and the need for flexible trading strategies.

The results emphasize the diverse techniques used by RL models in trading stocks. Although A2C, TD3, and PPO have shown promise for financial trading, further study is necessary to comprehend the underlying factors that contribute to their effectiveness. In addition, they must persist in improving their trading procedures for practical implementation. Our findings provide valuable insights into the use of RL in financial trading and emphasize the need of continuous research and improvement to enhance performance and adaptability in real-world scenarios.

\subsection{Purchase Diversity}
\begin{figure}[htbp]
    \centering
    \includegraphics[width=1\linewidth]{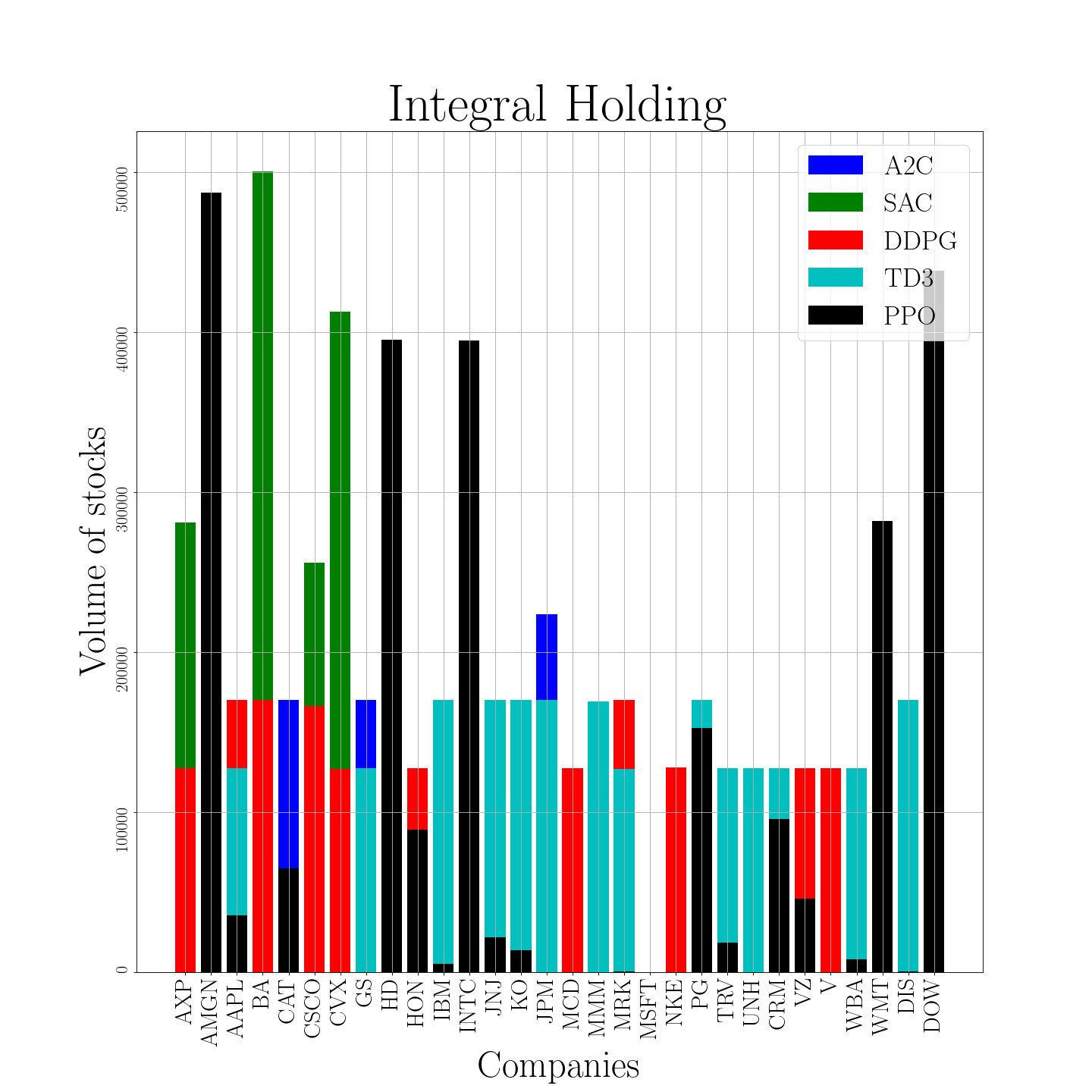}
    \caption{The integral holding values, which show the overall transaction volumes of certain stocks by each model for the whole testing period.
}
  \label{fig:integral_holding}
\end{figure}

As it is discussed, diversifying stock purchases is known to be paramount for investors seeking to mitigate risk and enhance long-term returns. Therefore, our next investigation helps to learn the behaviour of DRL algorithms to learn whether they tend to diversify their purchase behaviour.  With this objective in mind, we extract integral holding that is plotted in Figure \ref{fig:integral_holding}.  Integral Holding plot(Figure \ref{fig:integral_holding})offers a comprehensive analysis of the trading patterns of each RL model across many stocks. PPO used a varied approach, doing significant trades with some stocks while maintaining little activity with others. SAC focused its trading activities on a limited number of stocks and carried out large-scale transactions. A2C had comparable trading patterns to SAC, with a concentration on a limited selection of firms, but with lower trading volumes.

On the other hand, TD3 used a well-rounded approach by engaging in trading activities with a wide range of securities in very small amounts. DDPG had a similar pattern to TD3, but with less levels of trade activity.

\subsection{Holder vs Trader Algorithms}
\begin{figure}[htbp]
    \includegraphics[width=1\linewidth]{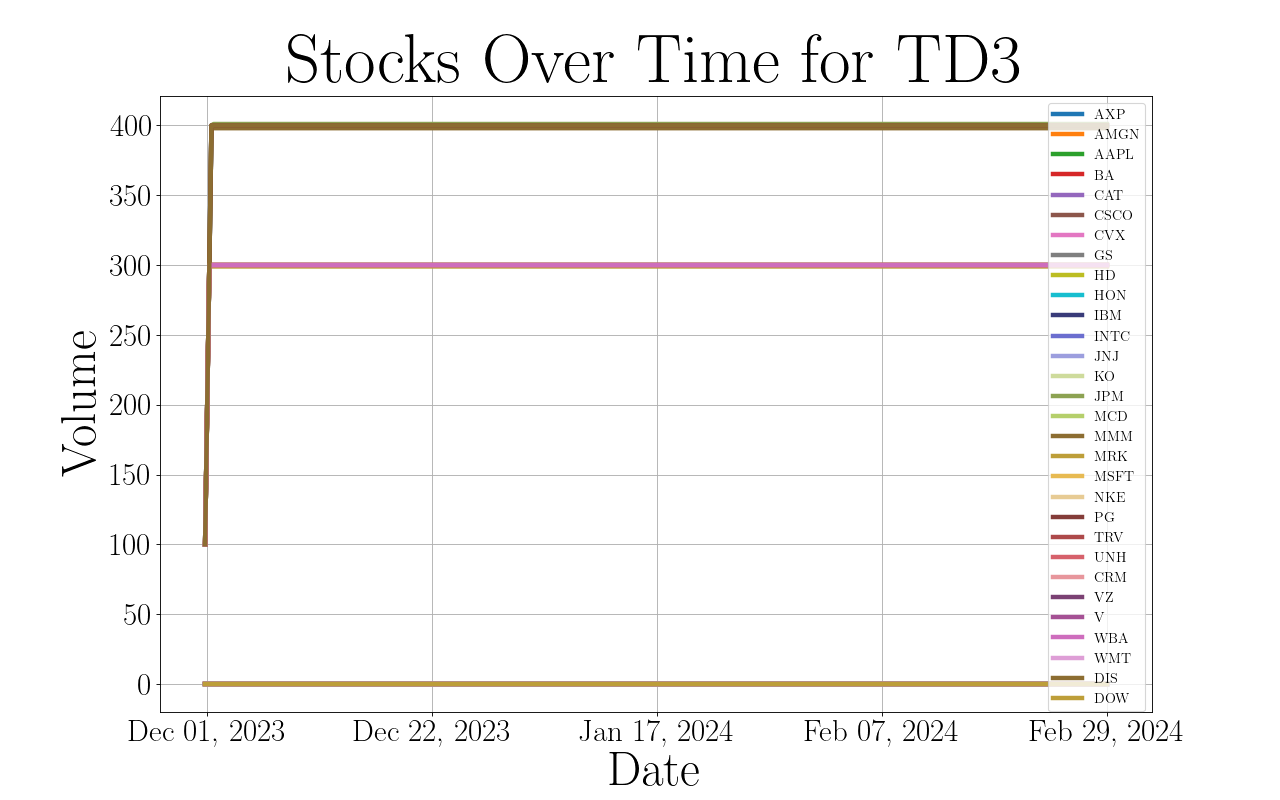}
    \caption{The stock holdings maintained by TD3 during the trading period.}
     \label{fig:td3}
\end{figure}

The figure \ref{fig:td3}, illustrate the trading patterns of TD3 RL model at different time periods. This figure demonstrated that TD3 tends to pick a few companies and hold them for the entire trading time. Although TD3 tends to hold only a few companies, figure \ref{fig:reward} shows that TD3 is one of the top performing DRL algorithms to to accumulate higher rewards. This means TD3 learn to hold companies to return competitively high reward.
The graph illustrates the maximum number of shares purchased by the Twin Delayed DDPG (TD3) algorithm, revealing a notable trend where the quantity of shares acquired consistently remains below 400. This pattern suggests a strategic limitation or preference within the TD3 algorithm, implying that it tends to avoid excessively large share acquisitions. Such behavior could be attributed to risk mitigation strategies embedded within the algorithm, aiming to maintain a balanced portfolio or prevent over-exposure to individual assets. Understanding this tendency is crucial for optimizing trading strategies and risk management protocols.

\begin{figure}[htbp]
    \includegraphics[width=1\linewidth]{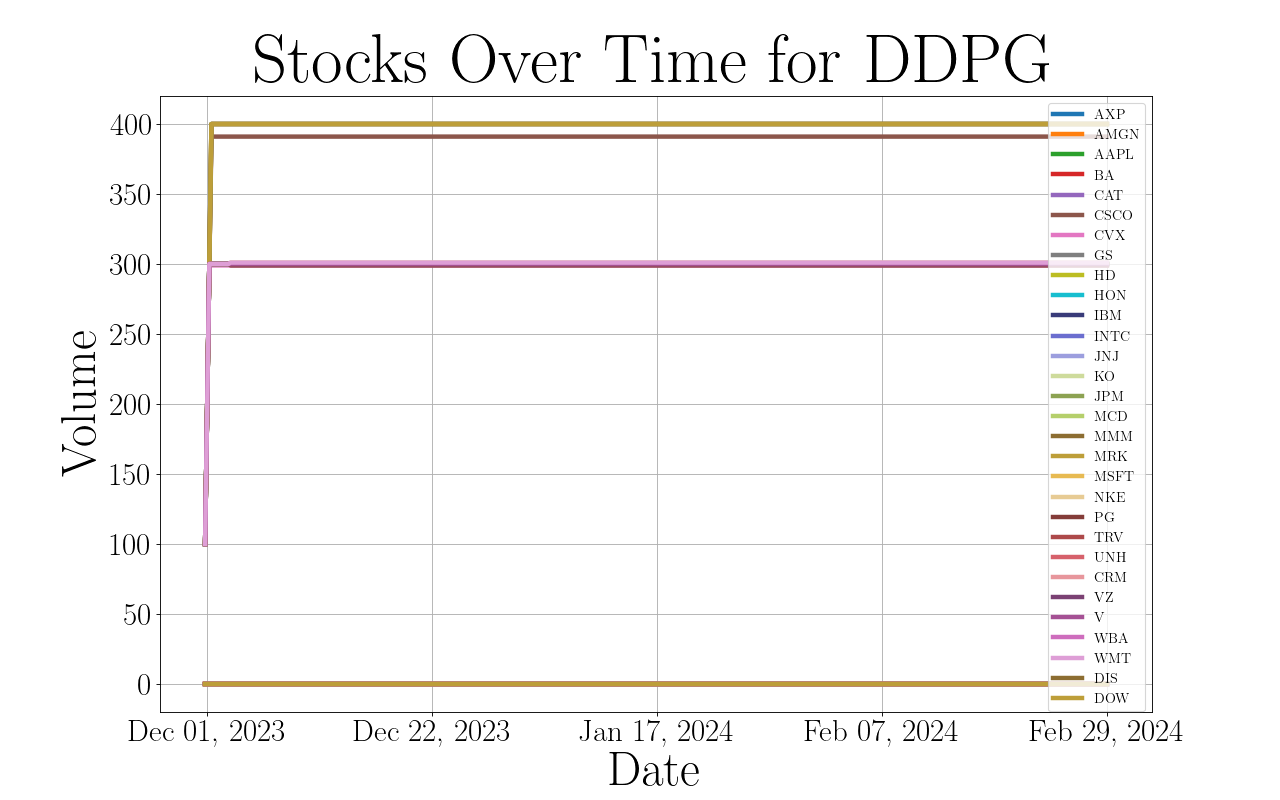}
    \caption{The stock holdings maintained by DDPG during the trading period. }
     \label{fig:DDPG}
\end{figure}

The figure \ref{fig:DDPG} illustrate the trading patterns of DDPG DRL model which shows the tendency of this model to holding instead of trading. Similar to Twin Delayed DDPG (TD3), DDPG algorithms exhibit a common inclination toward holding a limited number of shares, not exceeding 400. This shared characteristic reflects a deliberate strategy within both algorithms, possibly rooted in risk management principles or inherent biases in their learning processes. By maintaining a restrained portfolio size, these algorithms may aim to mitigate the potential downside risk associated with overexposure to individual assets or sectors. This preference for moderation underscores the algorithms' prudence in balancing potential rewards with the need to avoid excessive risk, highlighting their adaptive nature in navigating the complexities of financial markets. Understanding this tendency is integral to optimizing trading strategies and aligning investment decisions with the algorithms' inherent preferences for achieving robust and sustainable performance outcomes.

\begin{figure}[htbp]
    \includegraphics[width=1\linewidth]{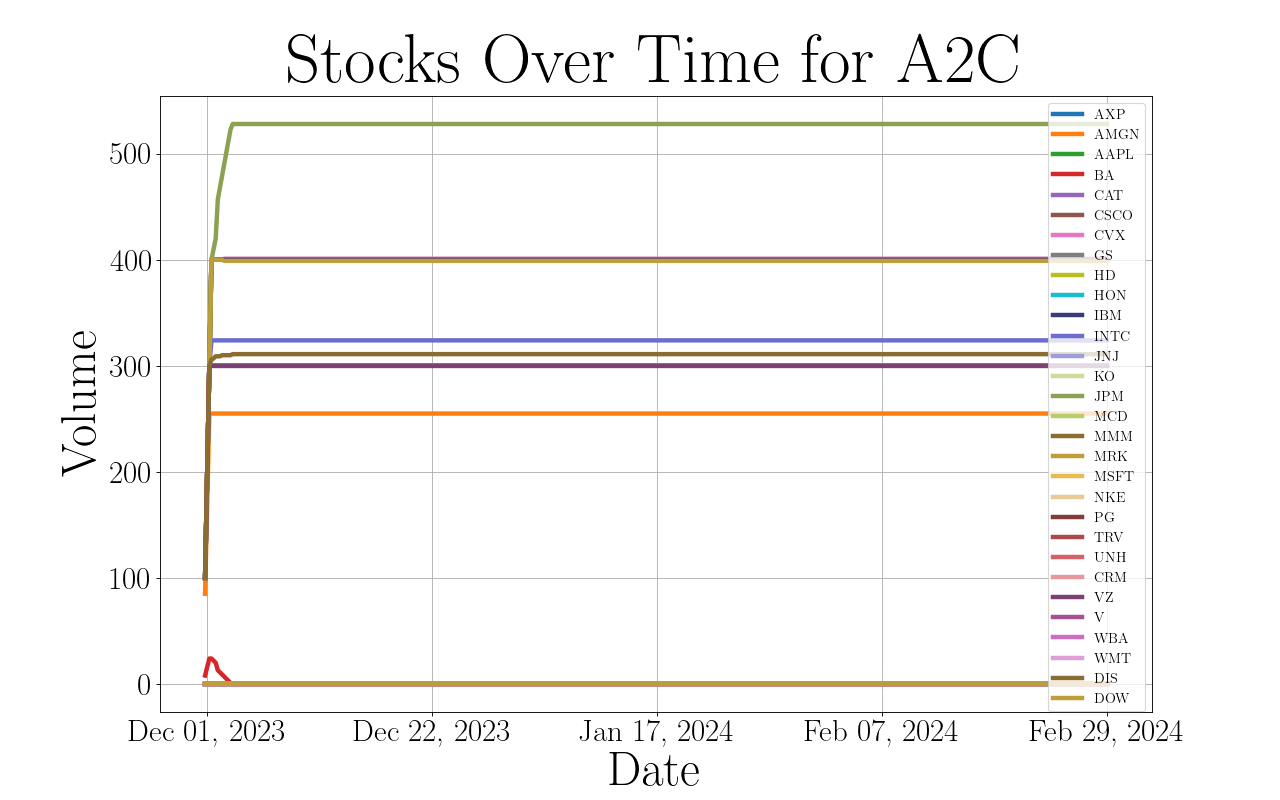}
    \caption{The stock holdings maintained by A2C during the trading period.}
     \label{fig:A2C}
\end{figure}

The figure \ref{fig:A2C} illustrate the trading patterns of A2C DRL model. While the A2C algorithm shares the inclination of holding a limited number of shares, akin to the DDPG and TTD3 algorithms, it distinguishes itself by incorporating a broader degree of diversity in its holdings. A2C tends to maintain a portfolio size of less than 500 shares, reflecting a similar risk-conscious approach to its counterparts. However, what sets A2C apart is its ability to introduce greater variety into its investment selections, potentially spanning across a wider range of stocks and sectors. This emphasis on diversity within its holdings enables A2C to capture a more number of companies while still adhering to its risk management objectives.

As oppose to first three algorithms (TD3, DDPG, A2C) the last two algorithms (PPO and SAC) tend to trade more than holding. The strategies of SAC and PPO shown a tendency for shorter duration of holding positions, often engaging in the actions of selling and purchasing stocks, as well as adjusting the amounts held. 

\begin{figure}[htbp]
    \includegraphics[width=1\linewidth]{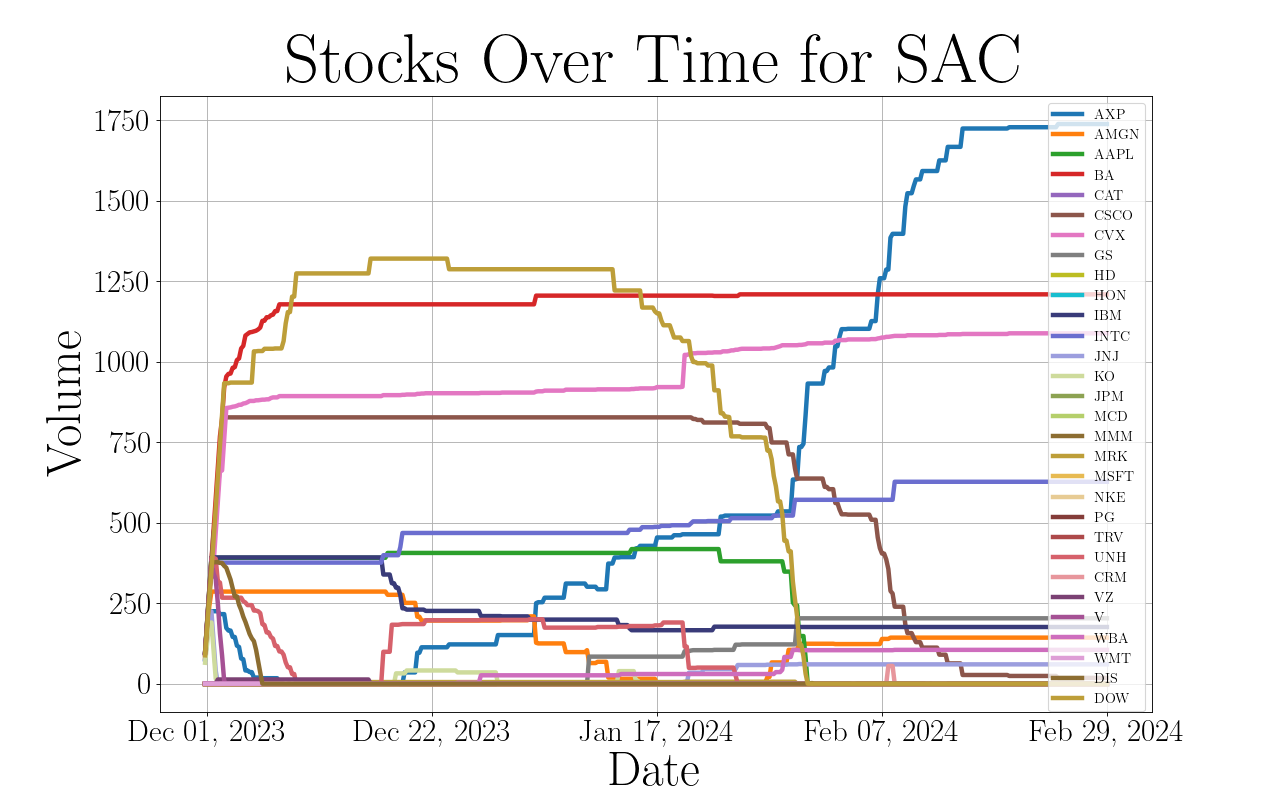}
    \caption{The stock holdings maintained by SAC during the trading period. }
     \label{fig:SAC}
\end{figure}

\begin{figure}[htbp]
    \includegraphics[width=1\linewidth]{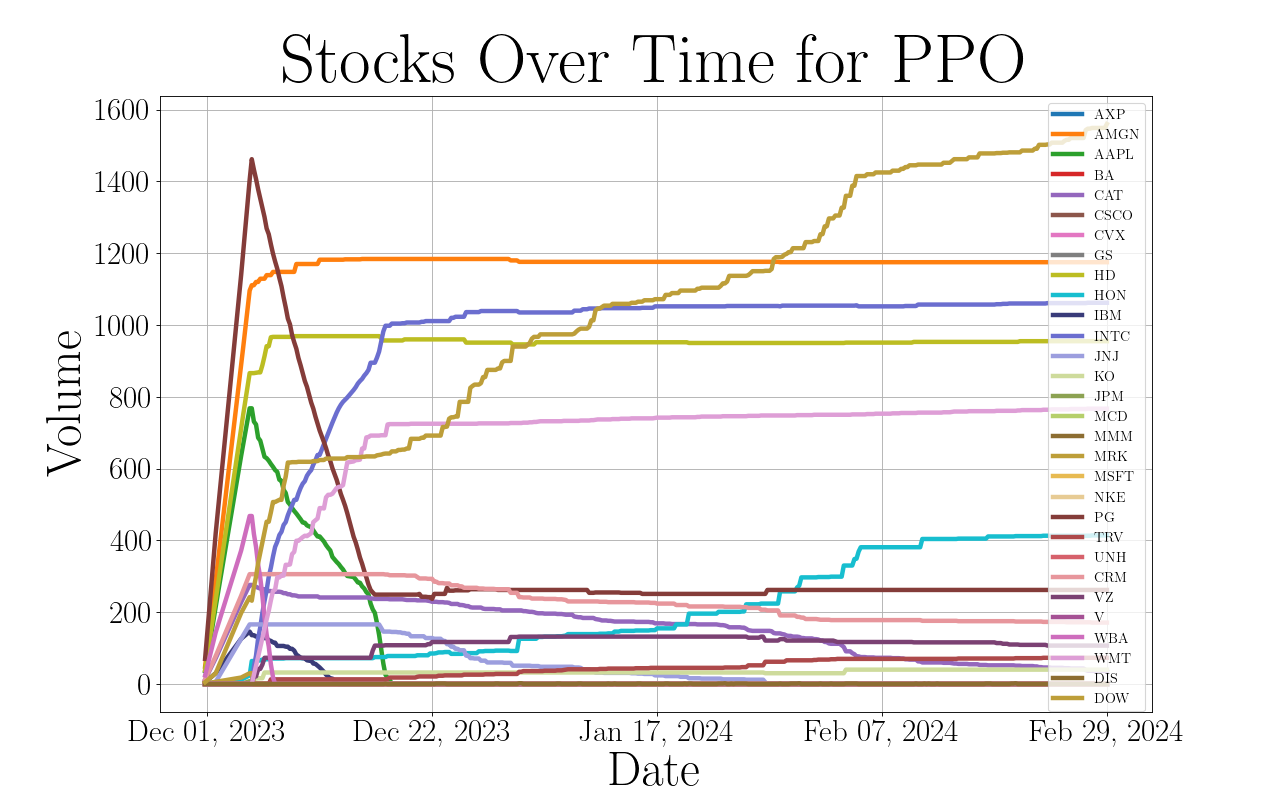}
    \caption{The stock holdings maintained by PPO during the trading period. }
     \label{fig:PPO}
\end{figure}

Figure \ref{fig:SAC}, and \ref{fig:PPO} demonstart the trading behaviour of PPO and SAC algoritthms. As it is shown , PPO and SAC tend to engage in a higher volume of trades, resulting in a significantly higher maximum purchase quantity. With maximum purchases around 1700 shares, these trader algorithms demonstrate a willingness to take more frequent and substantial positions in the market. In contrast, A2C, DDPG, and TD3 algorithms typically exhibit a more restrained trading behavior, with maximum purchases seldom exceeding 400 shares. This difference underscores the varying risk profiles and trading styles inherent in each algorithm, with PPO and SAC favoring a more dynamic and aggressive investment strategy compared to the relatively conservative nature of A2C, DDPG, and TD3.

\section{Conclusion}

In this research, we examined the performance and behavior of diverse Deep Reinforcement Learning (DRL) algorithms, including DDPG, PPO, TD3, SAC, and A2C, in the domain of financial trading. Our investigation aimed to gain insights into their decision-making processes, particularly concerning the critical choice between holding or trading assets. The results demonstrate that each algorithm exhibits unique trading patterns and strategies, with A2C emerging as the top performer in terms of cumulative rewards.

The analysis of the integral holding values reveals that while PPO and SAC engage in significant trades with a limited number of stocks, DDPG and TD3 adopt a more balanced approach, trading smaller amounts across a wider range of securities. Furthermore, the temporal evolution of stock holdings indicates that SAC and PPO tend to hold positions for shorter durations, frequently buying and selling stocks, whereas DDPG, A2C, and TD3 display a propensity to remain stationary for extended periods.

These findings underscore the complexity and dynamic nature of financial markets, emphasizing the necessity for adaptable and flexible trading strategies. Although the DRL algorithms showcased promising results, further research is essential to fully comprehend the underlying factors contributing to their performance and to continuously refine their trading methodologies for real-world implementation.

Moreover, our study highlights the potential of DRL in revolutionizing financial decision-making by uncovering hidden patterns, exploiting inefficiencies, and optimizing trading strategies in real-time. As DRL continues to advance, it is crucial to explore its implications for market efficiency, risk management, and investor returns.

In conclusion, this research provides valuable insights into the application of DRL in financial trading and underscores the importance of ongoing research and development to enhance the performance and adaptability of these algorithms in practical settings. By harnessing the power of DRL, we can pave the way for more intelligent, efficient, and profitable financial decision-making in the future.






%

\bibliographystyle{IEEEtran} 
\bibliography{trader_or_holder}


\end{document}